\documentclass[12pt]{article}
\pdfoutput = 1
\usepackage{graphicx}
\usepackage{subcaption}
\usepackage{titling}
\usepackage{authblk}
\usepackage{placeins}
\usepackage{color,soul}
\usepackage{multibbl}
\usepackage{hyperref}
\usepackage[space]{grffile} 
\graphicspath{{Figures/}}
\definecolor{Orchid}{rgb}{0.855, 0.439,0.839}
\definecolor{Moss}{rgb}{0.2, 0.8,0.2}

\makeatletter
\newcommand*\reset[1]{
	\renewcommand\AB@affillist{}%
	\global\let\AB@authlist\@empty
	\renewcommand\AB@authlist{}%
	\setcounter{affil}{0}%
	\setcounter{authors}{0}%
	\emptythanks
	}

\makeatother
\begin{document}

\newbibliography{main}
\bibliographystyle{main}{References/mycustombib}
\newbibliography{app}
\bibliographystyle{app}{References/mycustombibSI}

\date{}

\title{ Absorption Enhancement for Ultra-Thin Solar Fuel Devices with Plasmonic Gratings}

\author[1,2,3]{Phillip Manley\thanks{phillip.manley@helmholtz-berlin.de}}
\author[4]{Fatwa F. Abdi}
\author[4]{Sean Berglund}
\author[5]{A.T.M. Nazmul Islam}
\author[3]{Sven Burger}
\author[4]{Roel van de Krol}
\author[1,6]{Martina Schmid}

\affil[1]{Nanooptical Concepts for Photovoltaics, Helmholtz-Zentrum Berlin f\"{u}r Materialien und Energie GmbH, Hahn-Meitner-Platz 1, 14109 Berlin, Germany}
\affil[2]{Nanostructured Silicon for Photonic and Photovoltaic Implementations, Helmholtz-Zentrum Berlin f\"{u}r Materialien und Energie GmbH, Kekul\'{e}str. 5, 12489 Berlin, Germany}
\affil[3]{Zuse Institute Berlin, Takustr. 7, 14195 Berlin, Germany}
\affil[4]{Institute for Solar Fuels, Helmholtz-Zentrum Berlin f\"{u}r Materialien und Energie GmbH, Hahn-Meitner-Platz 1, 14109 Berlin, Germany}
\affil[5]{Institute for Quantum Phenomena in Novel Materials, Helmholtz-Zentrum Berlin f\"{u}r Materialien und Energie GmbH, Hahn-Meitner-Platz 1, 14109 Berlin, Germany}
\affil[6]{University of Duisburg-Essen and CENIDE, Lotharstr. 1, 47057 Duisburg, Germany}

\maketitle

This paper was published in ACS Applied Energy Materials \textbf{1} p.5810-5815 (2018) doi: \href{dx.doi.org/10.1021/acsaem.8b01070}{10.1021/acsaem.8b01070} and is made available as an electronic preprint with permission from the American Chemical Society.

\begin{abstract}
	


We present a concept for an ultra-thin solar fuel device with a nanostructured back contact. Using rigorous simulations we show that the nanostructuring significantly increases the absorption in the semiconductor, CuBi$_2$O$_4$ in this case, by 47\% (5.2~mAcm$^{-2}$) through the excitation of plasmonic modes. We are able to attribute the resonances in the device to metal-insulator-metal plasmons coupled to either localised surface plasmon resonances or surface plasmon polaritons.


Rounding applied to the metallic corners leads to a blueshift in the resonance wavelength while maintaining absorption enhancement, thus supporting the possibility for a successful realization of the device. For a 2D array, the tolerance of the polarization-dependent absorption enhancement is investigated and compared to a planar structure. The device maintains an absorption enhancement up to incident angles of 75$^{\circ}$. 

The study highlights the high potential for plasmonics in ultra-thin opto-electronic devices such as in solar fuel generation. 
\end{abstract}

\section{Introduction}
The conversion of sunlight to storable fuel is a challenge of paramount importance to modern society. Photoelectrochemical devices, consisting of semiconductor photoelectrodes immersed in aqueous solution, are a particularly interesting way to achieve this solar-to-fuel conversion. In recent years many different device designs based on various materials have seen intensive research \cite{main}{Montoya2017,Park2006}. Among these, metal oxides are particularly interesting since they possess general aqueous stability and are relatively inexpensive \cite{main}{Sivula2013,Abdi2013_2}. However, they share a common drawback, that of the discrepancy between carrier transport and light absorption. Due to the poor transport properties, the carrier diffusion length in oxides is typically less than 100 nm \cite{main}{Joly2006, Cherepy1998, Kennedy1978, Paracchino2012, Abdi2013, Berglund2016, Abdi2017}. On the other hand, they normally have an indirect band gap; relatively thick films ($>$500~nm) are needed to absorb enough light. This mismatch often severely limits the performance of a metal oxide photoelectrode.

In the present work we focus on the metal-oxide semiconductor CuBi$_2$O$_4$, which is an emerging p-type semiconductor for solar fuel applications. It has a band gap of around 1.8 eV, which is ideal for a top absorber in a tandem configuration \cite{main}{Arai2007, Hahn2012, Berglund2013, Berglund2016}. It also has a suitable band position; the conduction and valence band edges straddle both water reduction and oxidation potentials. As a result, the photocurrent onset potential for CuBi$_2$O$_4$ has been reported to be $\sim$1 V vs reversible hydrogen electrode (RHE), which is beneficial for a tandem configuration \cite{main}{Arai2007, Hahn2012, Berglund2013}. Previous reports of CuBi$_2$O$_4$ synthesis deposition have shown highly porous and irregular surface structures \cite{main}{Berglund2016,Hahn2012,Kang2016}. Alternative synthesis methods such as spray-pyrolysis have been used to make dense, homogeneous CuBi2O4 thin films \cite{main}{Wang2017} and new methods such as pulsed laser deposition (PLD) could be used to obtain a highly uniform CuBi2O4 ultra-thin film. PLD has been demonstrated as a viable option for other metal oxide photoelectrode materials including resonant light trapping Ti-doped alpha-Fe2O3 \cite{main}{Dotan2013}.

However, the material is limited by the optical absorption vs. carrier transport mismatch as was previously mentioned. The diffusion length is in the range of $\sim$50 nm, while a thickness of more than 500~nm is needed to absorb 90\% of the incident light \cite{main}{Berglund2016}. This is especially true for the longer wavelengths ($>$ 450~nm), where the quantum efficiency has been reported to be very low.

In order to combat this challenge, we propose to reduce the semiconductor thickness to 100~nm which should ensure efficient carrier collection by shortening the length photo-generated carriers have to travel. Simultaneously we use light management to obtain a sufficient absorption of incident sunlight.

Various light management strategies applied to the field of solar fuels have been presented in the literature. Photonic crystal structuring of the active layer has been used to enhance absorption \cite{main}{Jeremy2016}. Other approaches use dielectric particles and gratings to localize light inside the active layer \cite{main}{Kim2014,Cheng2018}.Structuring the active material into nanorods can also increase absorption while maintaining short carrier diffusion lengths \cite{main}{Pihosh2015}. A further proposed light management strategy is that of plasmonics \cite{main}{Thomann2011,Abdi2014}. Plasmonic metallic nanoparticles  act as optical antennas, allowing light to be concentrated in the vicinity of the semiconductor material, thereby enhancing absorption \cite{main}{Atwater2010,Schmid2016}. Furthermore metallic particles themselves may have beneficial catalytic properties \cite{main}{Berglund2013,Li2013}. Despite this, certain challenges are present for particles, such as quenching the photocatalysis process through recombination \cite{main}{DiVece2012,Govorov2006}.

In this paper we circumvent the challenges of using particles by considering a 100~nm thick CuBi$_2$O$_4$ layer on a grating consisting of laterally alternating Ag and SiO$_2$ on top of an Ag layer. Ag and SiO$_2$ are less positive than CuBi$_2$O$_4$ vs. RHE. In order to circumvent a Schottky at the rear interface, an additional back contact layer or heavy doping layer of the CuBi$_2$O$_4$ may be necessary \cite{main}{Hudait2001}. Since we focus on the optical device design, such a layer is not taken into account in the current work. The unit cell of this periodic structure is shown in the inset of figure \ref{fig:RTA}(b). We will refer to the SiO$_2$ region as a nanoslot since it forms a slot in the Ag. Plasmonic gratings have been realized for multiple applications including photovoltaic absorption enhancement \cite{main}{Paetzold2011} and biosensing \cite{main}{Iqbal2017}. For the current application, the nanoslot grating serves as the metallic back contact to transport photo-generated holes to the anode side as well as to enable better light management. A similar structure has been applied to infrared absorption enhancement in photovoltaics \cite{main}{Wang2013}. Through careful device design, we are able to shift the operational frequency to visible wavelengths.

\section{Results and Discussion}
\subsection{1D Grating}

\begin{figure}
	\begin{center}
		\includegraphics[width=\textwidth]{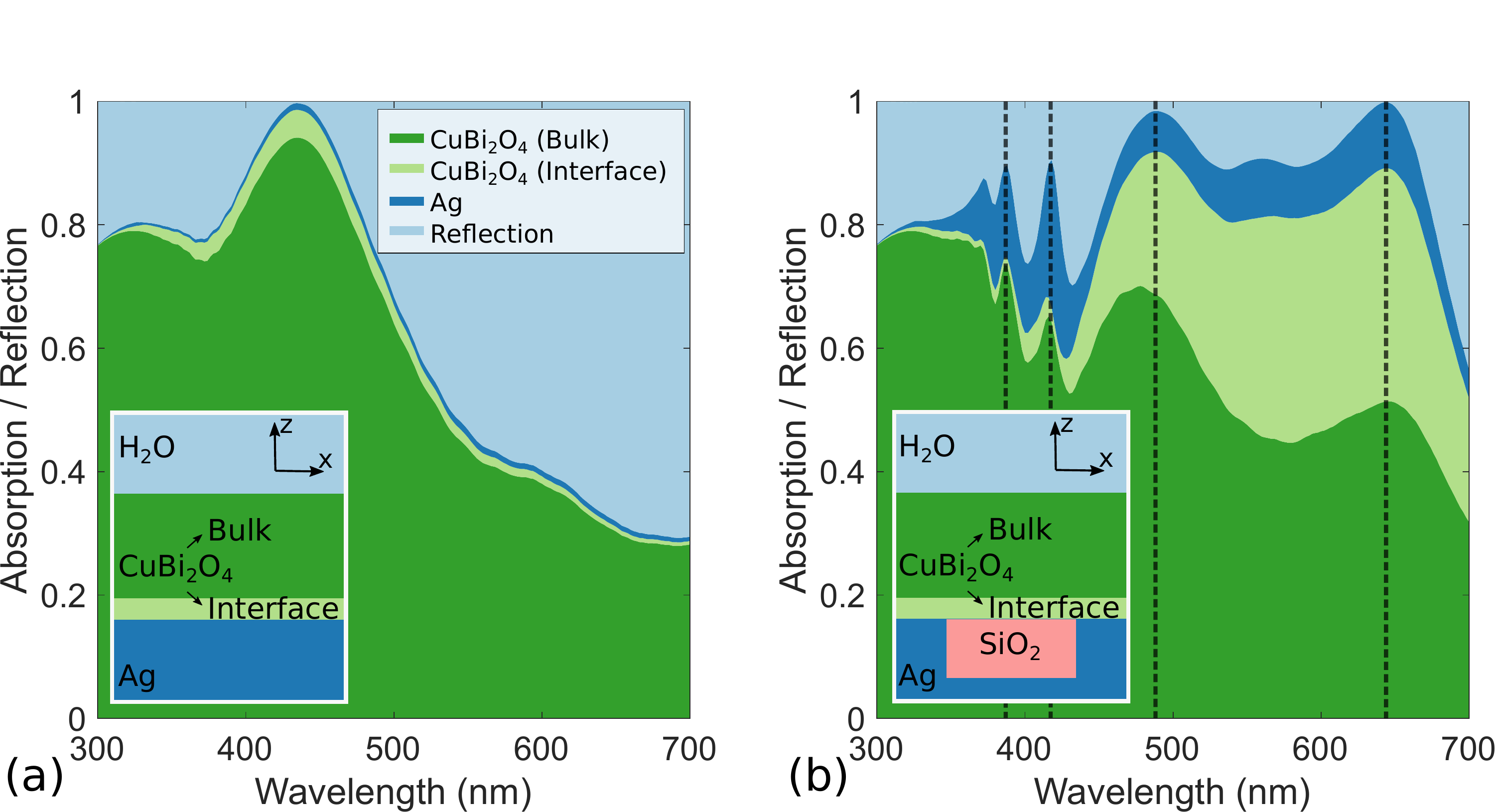}
	\end{center}

	\caption{The absorption in the 100 nm thick CuBi$_2$O$_4$ layer (divided into bulk [first 95~nm] and interface [last 5~nm] contributions), and the losses in Ag and Reflection for a solar fuel device with a flat Ag back contact (a) and a nanostructured Ag back contact (b). The dotted lines indicate resonance wavelengths shown in figure \ref{fig:NearFields}. Insets show schematic drawings of the periodic unit cell for each case.}
	\label{fig:RTA}
	
\end{figure}

Figure \ref{fig:RTA} shows the absorption and losses for the ultra-thin photoelectrode with a planar Ag back reflector with no nanoslots (a) and a nanostructured back reflector (b). The inset shows a schematic of each structure. Both of the devices have a 100~nm thick CuBi$_{2}$O$_{4}$ layer.

The catalytic reaction, in this case water reduction, occurs at the interface between H$_{2}$O and the inorganic semiconductor photocathode, CuBi$_{2}$O$_{4}$. Light is incident through the H$_{2}$O.

The nanostructured device has a unit cell width (pitch) of 112~nm, the SiO$_{2}$ slot is 70~nm wide and 60~nm deep, it is infinitely extended in the plane perpendicular to the page, therefore defining a 1D grating. These values were obtained from an optimization (figure \ref{fig:Opt}). We split the semiconductor into two regions. Firstly, the upper 95~nm of the semiconductor material (dark green), which is in contact with the water. Secondly, the final 5~nm of the semiconductor material (light green), which is in contact with the Ag back contact. These regions will be referred to as the 'bulk' and 'interface' regions respectively. Analogously, the absorption in each region will be referred to as 'bulk' and 'interface' absorption, respectively.

For the first case of the simple back reflector (no SiO$_2$ nanoslots), the interface absorption of the semiconductor is very small compared to the bulk. This is due to the  exponential damping of light while traversing the first 95~nm of the material (Lambert-Beer law) and also due to the volume of the region in question being much smaller than that of the bulk region. 

In order to quantify this absorption for solar fuel applications, we calculate the photocurrent density ($J_{abs}$) from the absorption curve and the solar spectrum. This assumes that all absorbed photons can contribute to water reduction. This assumption should be interpreted as the upper limit on photocurrent density for the proposed photoelectrode architecture. When modeling practical devices, the various loss mechanisms (e.g. recombination) have to be taken into account through the absorbed photon-to-current efficiency (APCE). We note that there have been reports for nanostructured metal-oxides with APCE values close to 100\% \cite{main}{Pihosh2015}. The proposed structure is able to reach a theoretical photocurrent density of 11.1 mAcm$^{-2}$ which is 54\% of the maximum achievable short circuit current density for a material with a band-gap of 1.8~eV (20.5 mAcm$^{-2}$).


Due to the low losses provided by Ag, the absorption in the back reflector is minimal, meaning that the main loss mechanism is reflection. This loss mechanism is eliminated entirely at the wavelength of 440~nm due to the presence of a Fabry-Perot resonance which eliminates the reflection. These kinds of resonances are clearly beneficial to absorption, however they cannot provide an arbitrarily broad absorption enhancement, since the only free parameter for tuning such a resonance is the film thickness.

Absorption is seemingly still present at wavelengths up to 700~nm. The complex refractive index used for CuBi$_{2}$O$_{4}$ also contains parasitic absorption at and below the band gap (figure \ref{fig:nk_data}). However analysis of the absorption coefficient has shown that the band gap for CuBi$_{2}$O$_{4}$ lies around 700~nm. Therefore we use this wavelength as the cutoff for absorption contributing to the photocurrent density. 


\begin{figure}
	\begin{center}
		\includegraphics[width=\textwidth]{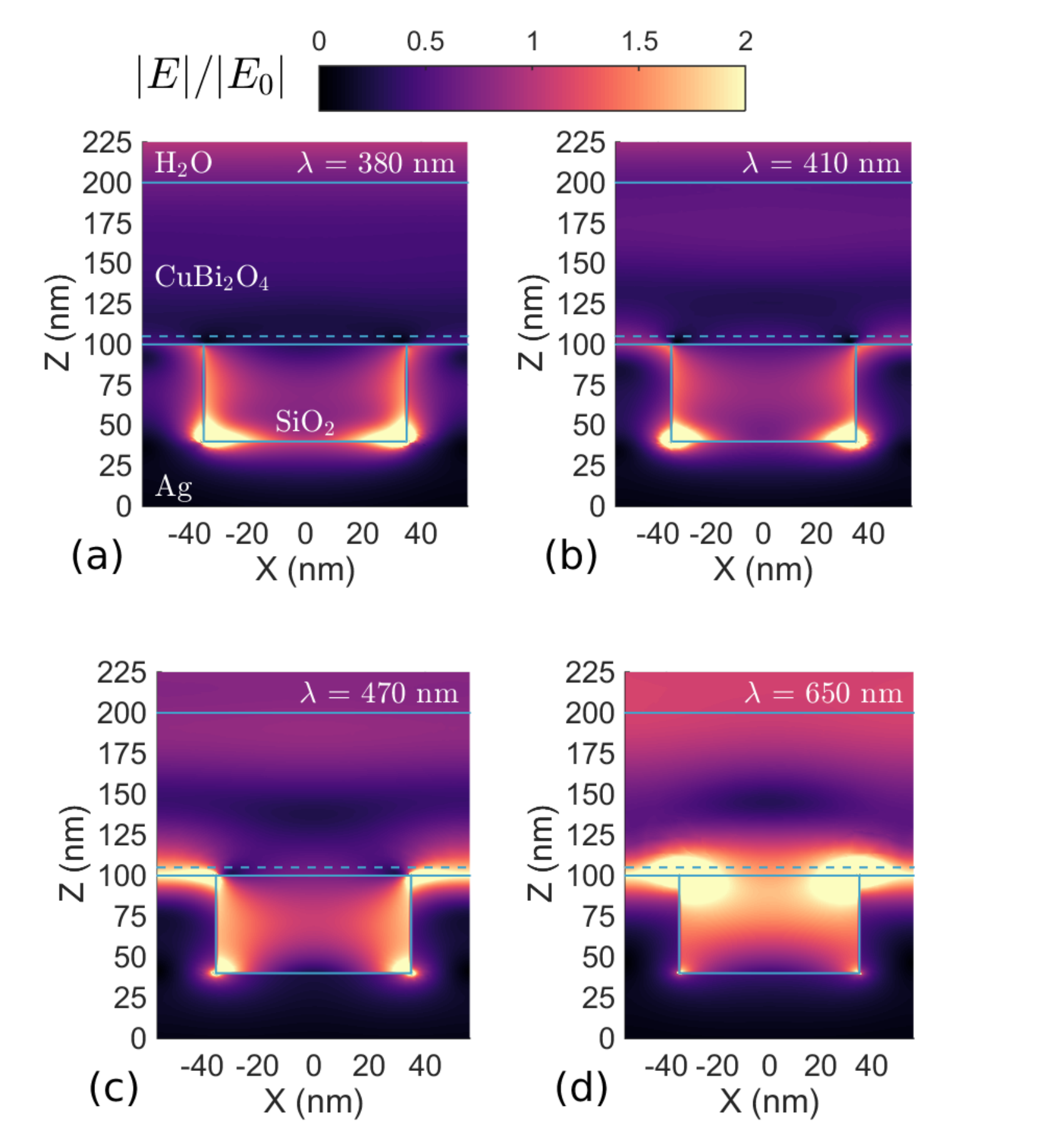}
	\end{center}
	
	\caption{The electric field in the nanostructured solar fuel device for four different wavelengths taken from the peaks in figure~\ref{fig:RTA}. Wavelengths are 380, 410, 470, 650~nm for parts (a-d), respectively. Light is incident normally from above and polarized in the x-z plane.}
	\label{fig:NearFields}	
\end{figure}

Figure \ref{fig:RTA}(b) shows the reflection and absorption for the nanostructured back contact (Ag with SiO$_2$ nanoslots). The polarization is oriented in the $x$~$-$~$z$ plane so that the plasmonic effects can be studied. In this case the total absorption is over 80\% for most of the visible spectrum, particularly between 500 and 650~nm where the solar spectrum has peak intensity. The overall absorption increase leads to a maximum short circuit current density of 16.3~mAcm$^{-2}$ (80\% of the maximum achievable). This large increase in absorption can be attributed to the excitation of resonant modes that exist at the Ag~/~CuBi$_2$O$_4$ interface, due to the large contribution of the interface absorption (light green). In this case the bulk absorption contributes 11.1~mAcm$^{-2}$ to $J_{abs}$ while the interface absorption contributes 5.2~mAcm$^{-2}$ to $J_{abs}$. In addition the parasitic absorption in the Ag back contact also increases due to increased interaction with the Ag arising from the resonant modes.


In order to investigate the mechanism of the absorption enhancement further, we show in figure \ref{fig:NearFields} the electric near field strength of the nanostructured back contact for the peak wavelengths shown in figure~\ref{fig:RTA}(b), namely 380, 410, 470 and 650~nm. The presence of a resonant mode is clearly shown in each case with strong localization of the electric field.

All four of the modes can be associated with metal-insulator-metal (MIM) plasmon resonances of the slot \cite{main}{Maier2007,Kurokawa2007}. This can be seen in the variations of the resonance wavelength with respect to the slot length which varies with a wavelength close to the analytical MIM mode wavelength (figures \ref{fig:Length_Variation} and \ref{fig:Length_Variation_Near_Fields}).

The four modes can be grouped into two categories. At the shorter wavelengths of 380 and 410~nm (figures \ref{fig:NearFields}(a) and (b)), the resonance is mainly confined to a localized mode at the bottom corners of the slot. In contrast, the resonances at the longer wavelengths of 470 and 650~nm are mainly located at the upper corners of the slot and at the Ag / CuBi$_2$O$_4$ interface.

An analysis of the variation of these modes with the device pitch (keeping the slot size constant) reveals that at wavelengths between 450 and 700~nm, the field localized at the upper corners acts as a source of surface plasmon polariton (SPP) modes (figures \ref{fig:Pitch_Variation} and \ref{fig:Pitch_Variation_Near_Fields}). Since the SPP modes have significant field strength in the CuBi$_2$O$_4$ they are able to increase the absorption there. In contrast the shorter wavelength modes remain localized in the bottom of the slot and therefore contribute mainly to losses in the Ag.

The maximum absorption enhancement is observed when an antinode of the MIM mode resonance is at the slot opening and when the SPP mode which is excited constructively interferes with itself. This holds true for the device at a wavelength of 650~nm. Slot length variations (figure \ref{fig:Length_Variation}) and pitch variations (figure \ref{fig:Pitch_Variation}) which affect the MIM and SPP resonances, respectively, show a peak for 60~nm slot length and 112~nm pitch. For the peak at 470~nm wavelength, the MIM mode is at maximum enhancement while the SPP is slightly off resonance. Conversely, at 575~nm wavelength, the MIM mode is off resonance while the SPP mode shows maximum constructive interference causing the increase in interface absorption visible at this wavelength.

Due to the coupling to SPP interface modes, the total absorber layer thickness can, in principle, be reduced while maintaining a high absorption. However, the total absorption obtained will still drop with decreasing layer thickness, as the short wavelength light ($<$ 350~nm) still needs to be absorbed conventionally since the inter-band transition losses in Ag prevent any beneficial resonances from forming at these wavelengths. Therefore, although we have presented absorption curves for the device with a 100~nm thick absorbing layer, we stress that the presented nanostructured device could also conceivably provide absorption enhancements for much thinner layers.

\subsection{Effect of Corner Rounding}

A further consideration for the implementation of these nanostructures in realistic devices is the ability to fabricate precise geometries. As the near field pictures from figure \ref{fig:NearFields} show, there is a strong localization of electric field in the vicinity of the sharp edges of the Ag grating. Such perfectly sharp edges are difficult to fabricate, therefore a certain amount of rounding on the edges should be expected.

\begin{figure}
	\begin{center}
		\includegraphics[width=0.5\textwidth]{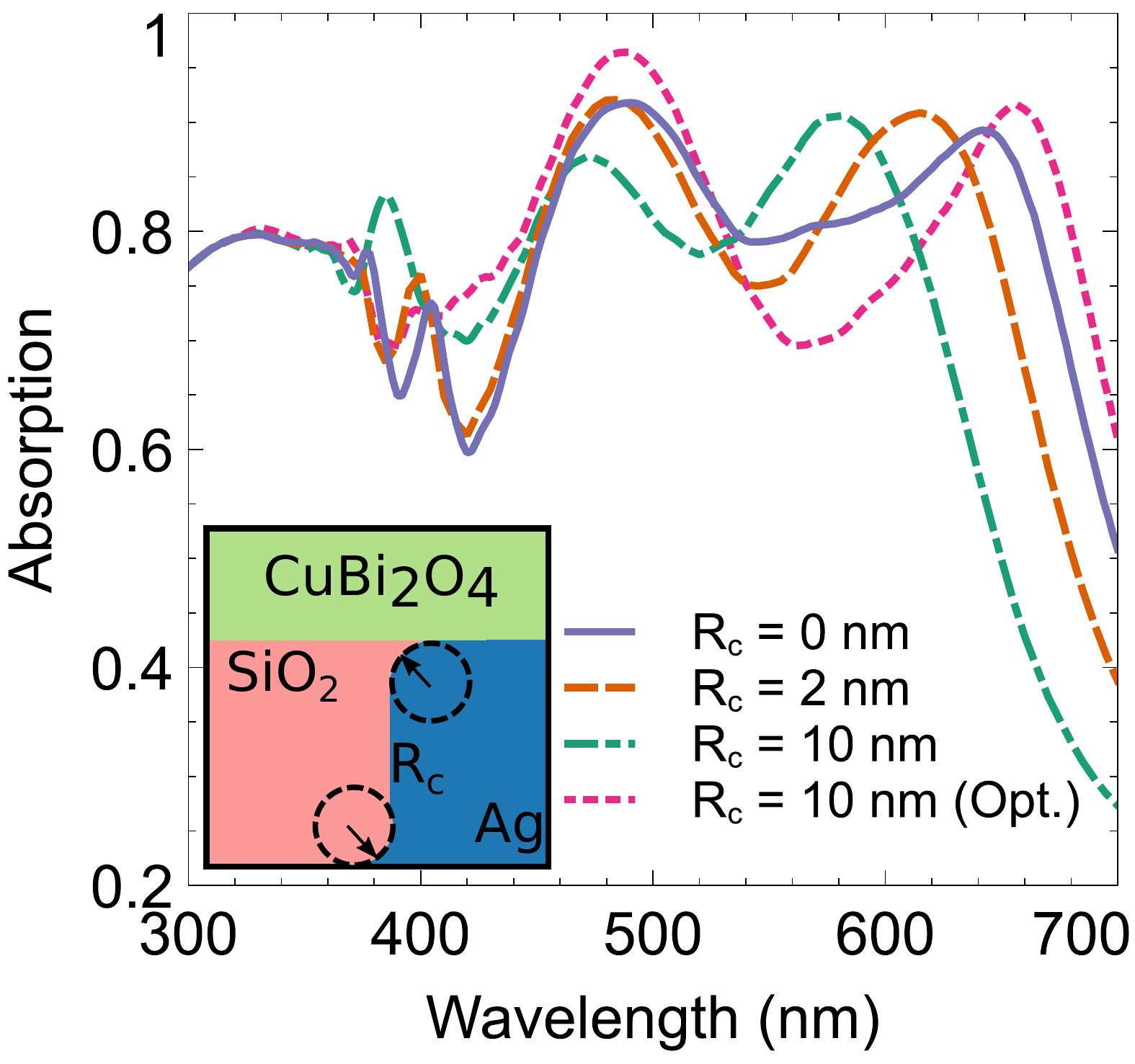}
	\end{center}
	
	\caption{The absorption inside the absorbing semiconductor (CuBi$_2$O$_4$) as a function of the wavelength for three different values of corner rounding $R_{c}$. The first three structures have the same geometrical parameters as in figure \ref{fig:RTA}(b), while the fourth has the geometrical parameters optimized for the corner rounding. Inset shows the definition of $R_{c}$. All other aspects of the geometry are the same as in figure \ref{fig:RTA}(b). Light is incident normally from above and polarized in the $x$-$z$ plane.}
	\label{fig:CornerRounding}
	
\end{figure}

In figure \ref{fig:CornerRounding} we show the absorption in the 100 nm of CuBi$_{2}$O$_{4}$ for the nanostructured 1D grating for three cases of corner rounding radius ($R_{c}$) at both the upper and lower grating corners. The definition of $R_{c}$ is shown in the inset of figure \ref{fig:CornerRounding}. The values of R$_{c}$ presented are 0 (same absorption curve as shown in figure~\ref{fig:RTA}(b)), 2~nm and 10~nm. It can be seen that even for $R_{c}$~=~2~nm, the resonances are blueshifted and this becomes more pronounced with increasing corner rounding.

The blueshift seen is a combination of multiple factors. The MIM mode resonance wavelength tends to decrease with decreasing slot width. This may be more relevant to the shorter wavelength resonances since they are localised to the bottom of the slot. Furthermore, due to the inhomogeneous width of the slot, the length of slot which has the necessary width for the supporting the MIM resonance will be effectively shorter. Since the MIM mode provides an absorption enhancement when an antinode is present at the slot opening, the wavelength of MIM mode necessary for this may be shifted to shorter wavelengths. Finally, as the corner rounding increases, more of the SPP resonance will be located inside the slot which has a lower refractive index than CuBi$_{2}$O$_{4}$. This lower refractive index will tend to redshift the SPP resonance. Due to competing factors, the effect of corner rounding on the exact resonance position is difficult to predict, necessitating further numerical study for an optimum to be found.




Despite the blueshifting the core resonant modes are all still present. We can conclude that the resonant absorption enhancement is not reliant on an unphysical singularity at the corners. This is important for the physical realization of the device.

The maximum photocurrent density obtained for the device with no corner rounding was previously shown to be 16.3~mAcm$^{-2}$. When a corner rounding of 2~nm is imposed, the photocurrent density lowers slightly to 15.8~mAcm$^{-2}$. As the corner rounding is increased to 10~nm, the photocurrent density further decreases to 14.6~mAcm$^{-2}$. All of these values show a significant improvement over the planar value of photocurrent density of 11.1~mAcm$^{-2}$. It should be further noted that the geometry can be reoptimized with corner rounding. If the corner rounding is 10~nm then a new optimum absorption enhancement can be found for a pitch of 200~nm, slot width of 100~nm and slot length of 50~nm. In this case the photocurrent density is increased to 16.5~mAcm$^{-2}$. The absorption profile for the optimized structure with corner rounding is shown in figure \ref{fig:CornerRounding}.

\subsection{2D Grating}

For integration into a solar fuel device, the proposed nanostructured grating has to continue to provide a strong enhancement in the presence of unpolarized light at different angles of incidence. In order to enhance the unpolarized response, the 1D grating can be extended to a 2D grating while keeping the same dimensions as the 1D grating. The resonance conditions found previously can be maintained for either a Ag grating with SiO$_{2}$ nanoslots (1), or a SiO$_{2}$ matrix containing Ag cubic nanoparticles connected to a Ag back contact (2). A schematic of configuration (2) is shown in figure \ref{fig:2DSchema}. Configuration (2) was found to be more optically beneficial than configuration (1) and was therefore chosen for the results presented in this section. For reasons of computational efficiency, corner rounding has not been used for the 2D grating.

\begin{figure}
	
	\centering
	\includegraphics[width=\textwidth]{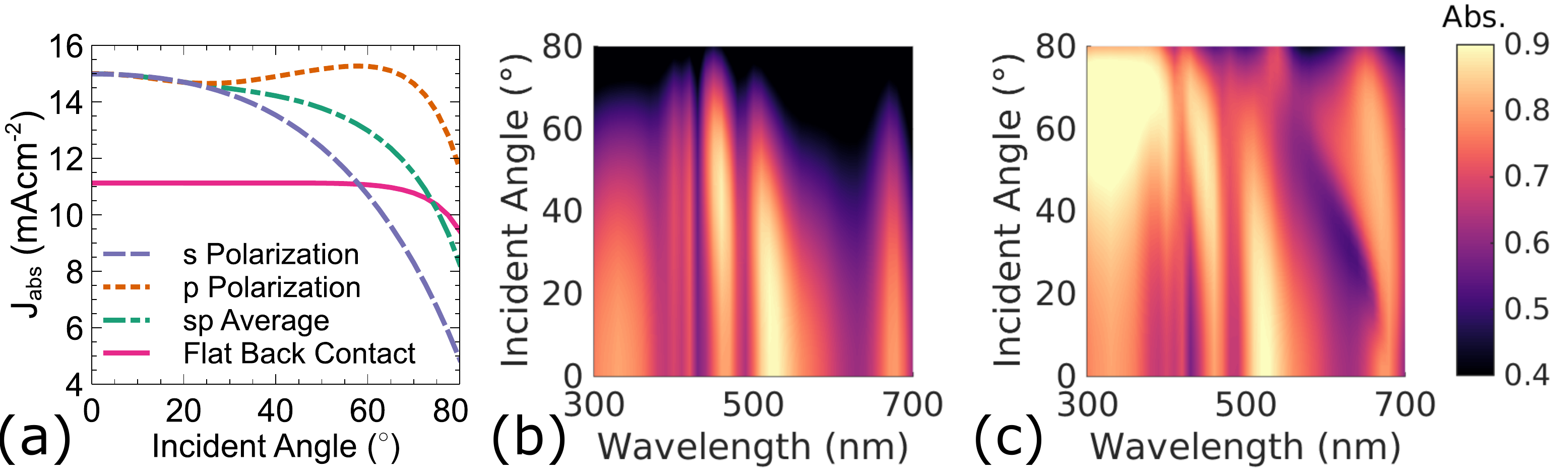}
	\caption{The absorbed photocurrent density as a function of the incident polar angle $\theta$ (a). The absorption in the semiconductor (CuBi$_2$O$_4$) for a 2D grating for polarization s (b) and p (c) as a function of both wavelength and polar incidence angle.}
	\label{fig:2DGrating}	
	
\end{figure}



The periodic unit cell of the 2D grating has a 3 fold mirror symmetry. The mirror planes lie along the $x$ axis, the $y$ axis and at 45$^{\circ}$ to the $x$ and $y$ axes. Therefore it is sufficient to use azimuthal angles $0^{\circ}<\phi<45^{\circ}$ to obtain the azimuthally averaged response. We chose 5 azimuthal angles equally spaced between 0$^{\circ}$ and 45$^{\circ}$. The difference in absorption between averaging over 3 and 5 angles was smaller than $10^{-3}$, therefore it was concluded that 5 angles are sufficient to obtain an averaged response. For each polar angle ($\theta$ in figure \ref{fig:2DSchema}) of incidence $> 0$ we can define two orthogonal polarizations: p polarization where the electric field orientation lies in the scattering plane and s polarization where the electric field orientation is perpendicular to the scattering plane, as shown in figure \ref{fig:2DSchema}. 


Figure \ref{fig:2DGrating}(a) shows how the maximum photocurrent density $J_{abs}$ for the structure depends on the incident polar angle. For oblique incidence p polarization provides a higher current density than for s polarization. Taking the unpolarized response into account, the grating outperforms a planar stack for angles up to 75$^{\circ}$ which is highly beneficial to solar fuel applications.

Figure \ref{fig:2DGrating}(b-c) shows the wavelength resolved polar angular response for s and p polarization. The 2D grating, even at normal incidence, does not show the same behavior as the 1D grating. This is to be expected as they are physically different systems, however the main resonances present in figure~\ref{fig:RTA}(b) are also present in figure~\ref{fig:2DGrating}, namely a strong absorption enhancement between 500 and 600~nm and a slightly weaker enhancement between 600 and 700~nm. In general the absorption for the case of p polarization is higher as we move towards higher angles. This is partly due to reflection at the initial H$_{2}$O/CuBi$_{2}$O$_{4}$ interface being lower for p polarization.

The dependence of the plasmonic resonances on incident angle differs for the p and s polarizations. In figure \ref{fig:2DGrating}(b) we see that for s polarization, the  resonance wavelengths remain constant with increasing angle. The resonance positions do not change due to the electric field remaining normal to the vertical sides of the Ag cuboids for all incident angles. For the case of p polarization shown in figure \ref{fig:2DGrating}(c), the resonances broaden and increase for higher angles. This is due to electric field component no longer being purely normal to the vertical sides of the Ag cuboids, thereby changing the resonance condition for excitation of MIM modes. Broader resonances will be beneficial to the broadband functioning of the device.



\section{Conclusion}
We have presented a nanostructured back contact for use in ultra-thin solar fuel devices. The nanostructuring was shown to significantly increase the absorption in the absorbing semiconductor, CuBi$_{2}$O$_{4}$ in this case, by 47\% (5.2 mAcm$^{-2}$) through the excitation of plasmonic modes.

By varying the length of the SiO$_2$ nanoslot and the device pitch the resonances could be classified as MIM modes which remain isolated in the nanoslot or couple to SPP modes dependent on the wavelength.

By simulating the effect of corner rounding, it could be confirmed that the presented results do not rely on unphysical singularities at material interfaces. This means that translation of the simulated results to an experimental reality is promising. Moreover, we demonstrated that the detrimental effect of corner rounding, which is unavoidable in practical systems, can be fully compensated by adjusting the pitch and slot dimensions of the SiO$_2$.

For a 2D grating based on the optimal 1D nanoslot grating, the angular tolerance of the absorption enhancement was investigated for two different polarizations. A greater angular tolerance was shown for p polarization, with the absorption even increasing at higher angles. The unpolarized response was shown to outperform planar layers up to angles of 75$^{\circ}$.

These results provide a clear pathway to overcome the mismatch between the optical absorption and carrier diffusion length, which is present in many semiconducting photoelectrodes. In the end, it is expected that higher solar-to-fuel conversion efficiency with metal oxide photoelectrodes, especially the promising photocathode material CuBi$_{2}$O$_{4}$, is to be achieved with our proposed architecture.

\section{Methods}
All simulations were done using the commercial software JCMsuite \cite{main}{Pomplun2007}, a finite element solver for Maxwell's equations. All simulations modelled the upper and lower half spaces with perfectly matched layers while periodic boundary conditions were used in the $x-y$ plane. A plane wave source was incident from the upper half space. For H$_2$O and SiO$_2$ a wavelength independent refractive index of 1.33 and 1.5 was used, respectively. For Ag the data of Johnson and Christy were used \cite{main}{Johnson1972}. The complex refractive index of CuBi$_2$O$_4$ was obtained via spectroscopic ellipsometry of a single crystal using an M-2000D ellipsometer (193-1000 nm, J.A. Woollam Co., Inc.). Details of crystal synthesis are given in the supporting information. By using a finite element degree of 3 and mesh side length constraint smaller or equal to one tenth of the wavelength in the material (for dielectrics) or a constant value of 5 nm (for metal) we were able to ensure an accuracy of greater than $10^{-3}$ for the absorption and reflection.

\section{ASSOCIATED CONTENT}

Supporting Information Available: The complex refractive index of CuBi$_{2}$O$_{4}$ used in all simulations. Description of the procedure for obtaining absorption and photocurrent density. The effects of length and pitch variations of the nanoslot. Optimization of the nanoslot with respect to slot width, slot length and pitch ratio. The geometry of the 2D nanoslot array and the definition of the scattering angles.

\section{Acknowledgements}
The authors would like to thank Dr. A. Bronneberg for ellipsometry measurements. P. Manley and M. Schmid would like to acknowledge funding and support from the Initiative and Networking fund of the Helmholtz Association for the Young Investigator Group VH-NG-928. Part of the work was done at the Berlin Joint Lab for Optical Simulations for Energy Research (BerOSE). P. Manley acknowledges funding from the Helmholtz Innovation Lab HySPRINT, which is financially supported by the Helmholtz Association.
\bibliography{main}{references}{References}

\makeatletter
\newcounter{mybibitem}
\def\bibitem#1{\stepcounter{mybibitem}\@lbibitem[S\the\value{mybibitem}]{#1}}
\makeatother

\newpage
\appendix
\reset{1}

\title{Supporting Information: Absorption Enhancement for Ultrathin Solar Fuel Devices with Plasmonic Gratings}
\author{}

{\maketitle}

\pagenumbering{arabic}
\renewcommand*{\thepage}{S-\arabic{page}}
\renewcommand\thefigure{S\arabic{figure}}    
\renewcommand\thetable{S\arabic{table}}    
\FloatBarrier
\setcounter{figure}{0}    
\label{sec:SupportingMaterial}

\begin{figure}
	\begin{center}
		\includegraphics[width=0.8\textwidth]{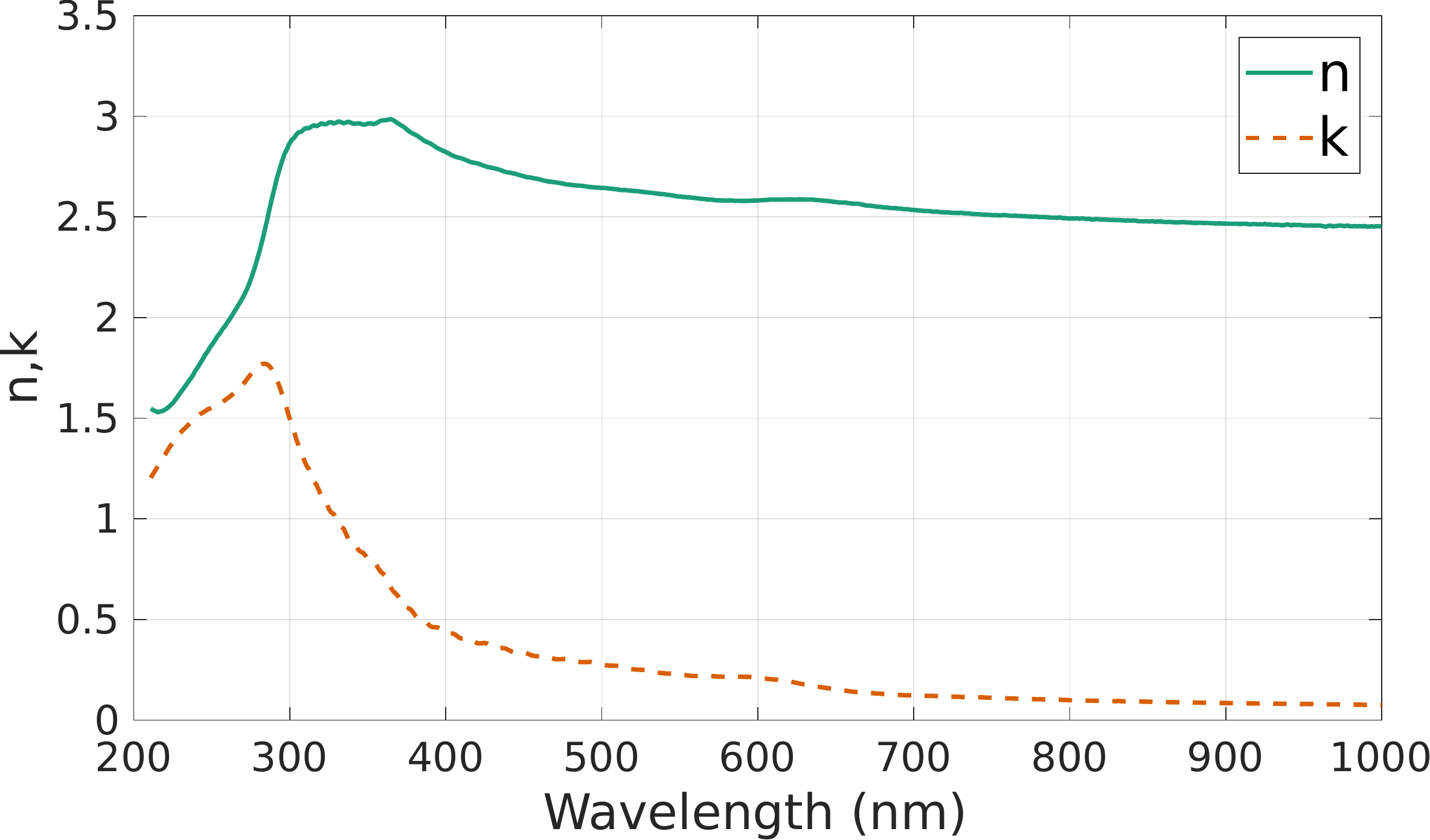}
	\end{center}	
	\caption{The real (n) and imaginary (k) parts of the refractive index of CuBi$_{2}$O$_{4}$ measured using spectroscopic ellipsometry.}
	\label{fig:nk_data}
\end{figure}

Figure \ref{fig:nk_data} shows the real (n) and imaginary (k) parts of the refractive index of CuBi$_{2}$O$_{4}$. The data were obtained from a spectroscopic ellipsometry measurements of a single crystal using an M-2000D ellipsometer (193-1000 nm, J.A. Woollam Co., Inc.). The crystal growth of CuBi$_{2}$O$_{4}$ was done by the floating zone technique in a four mirror type optical image furnace (Crystal Systems Corp., Japan) in the quantum materials corelabs at the Helmholtz-Zentrum Berlin. For crystal growth, high density feed rod (D= 6~mm, L= 7~cm) was prepared from a stoichiometric mixture of high purity of Bi$_{2}$O$_{3}$ (99.99\%) and CuO (99.995\%) by solid state reactions. Growth in the floating-zone machine was done in ambient air atmosphere at a rate of 5~mm/h.

After solving Maxwell's equations to obtain the electric field over the whole volume of our computational domain we wish to calculate related quantities. To obtain the absorption in each material, we use the imaginary part of the electric field energy density integrated over the volume of the material \cite{app}{Jackson1998},

\begin{equation}
A = 2\omega\int_{V_{m}} \epsilon_{m}'' |\vec{E}(\vec{r})|^{2}d\vec{r}
\end{equation}
Where $\omega$ is the frequency of the electric field, $\epsilon_{m}''$ is the imaginary part of the permittivity in a material $m$, $V_{m}$ is the volume of material $m$ and $\vec{E}(\vec{r})$ is the electric field.

The associated photocurrent density may be obtained by integrating the convolution of the wavelength dependent absorption with the number of photons per wavelength from the solar flux,

\begin{equation}
J_{abs} = -e \int_{0}^{\lambda_{g}} A(\lambda) \Phi(\lambda) \frac{\lambda}{hc}d\lambda
\end{equation}

Where $e$ is the electronic charge, $\lambda_{g}$ is the band gap of the material, $A(\lambda)$ is the wavelength dependent absorption, $\Phi(\lambda)$ is the solar flux, $h$ is Planck's constant and $c$ is the speed of light in vacuo. 

\begin{figure}
	\begin{center}
		\includegraphics[width=\textwidth]{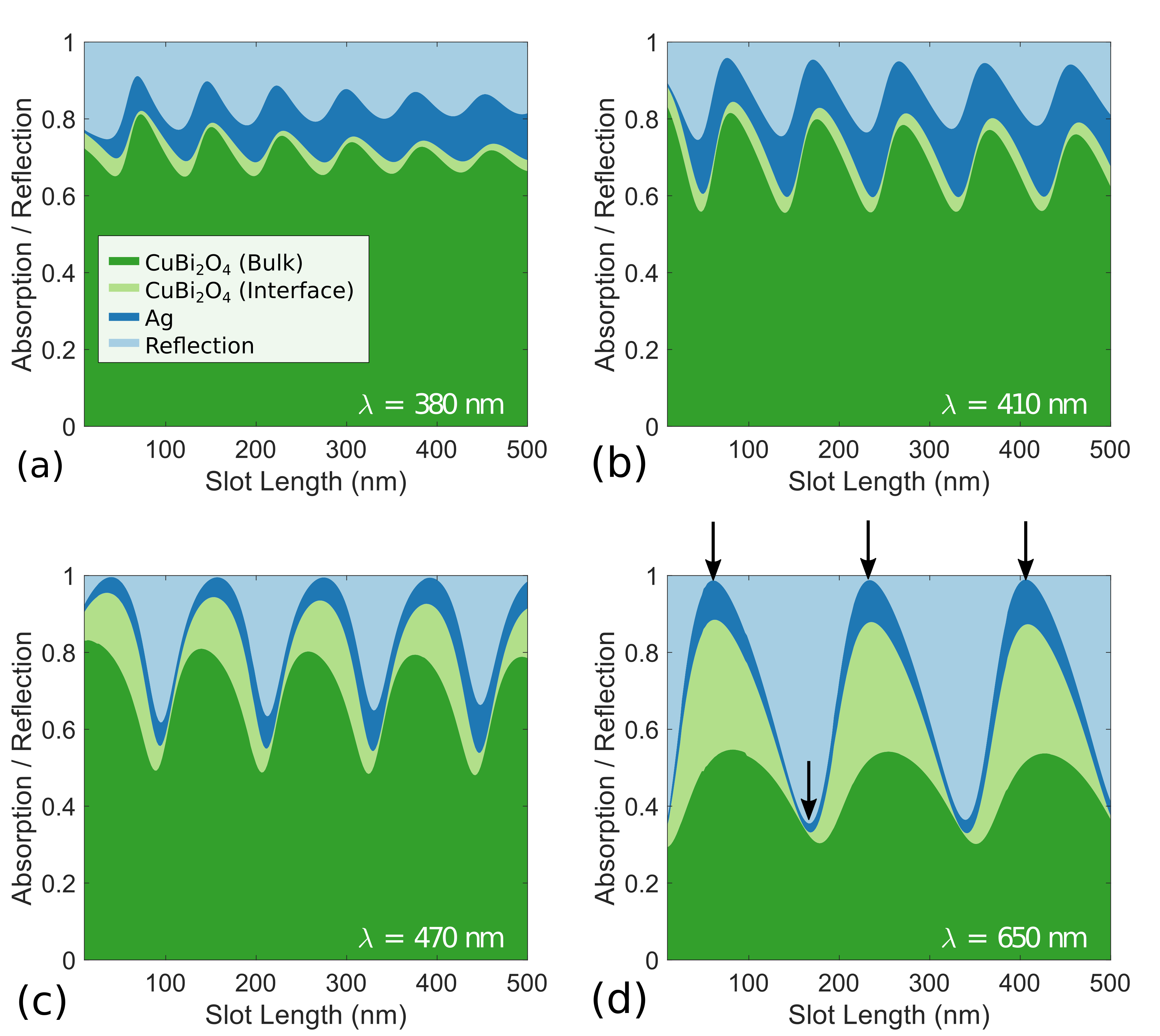}
	\end{center}	
	\caption{The absorption and reflection for the nanostructured 1D grating as a function of the SiO$_2$ slot length. The wavelength of light is 380, 410, 470, 650~nm for (a-d), respectively. The pitch and slot width are 112 and 70~nm, respectively. Arrows in (d) represent lengths for which the electric field data is shown in figure \ref{fig:Length_Variation_Near_Fields}. Apart from slot length the geometry is the same as in the inset of figure \ref{fig:RTA}(b).}
	\label{fig:Length_Variation}
\end{figure}

\begin{figure}
	\centering
	\includegraphics[width=\textwidth]{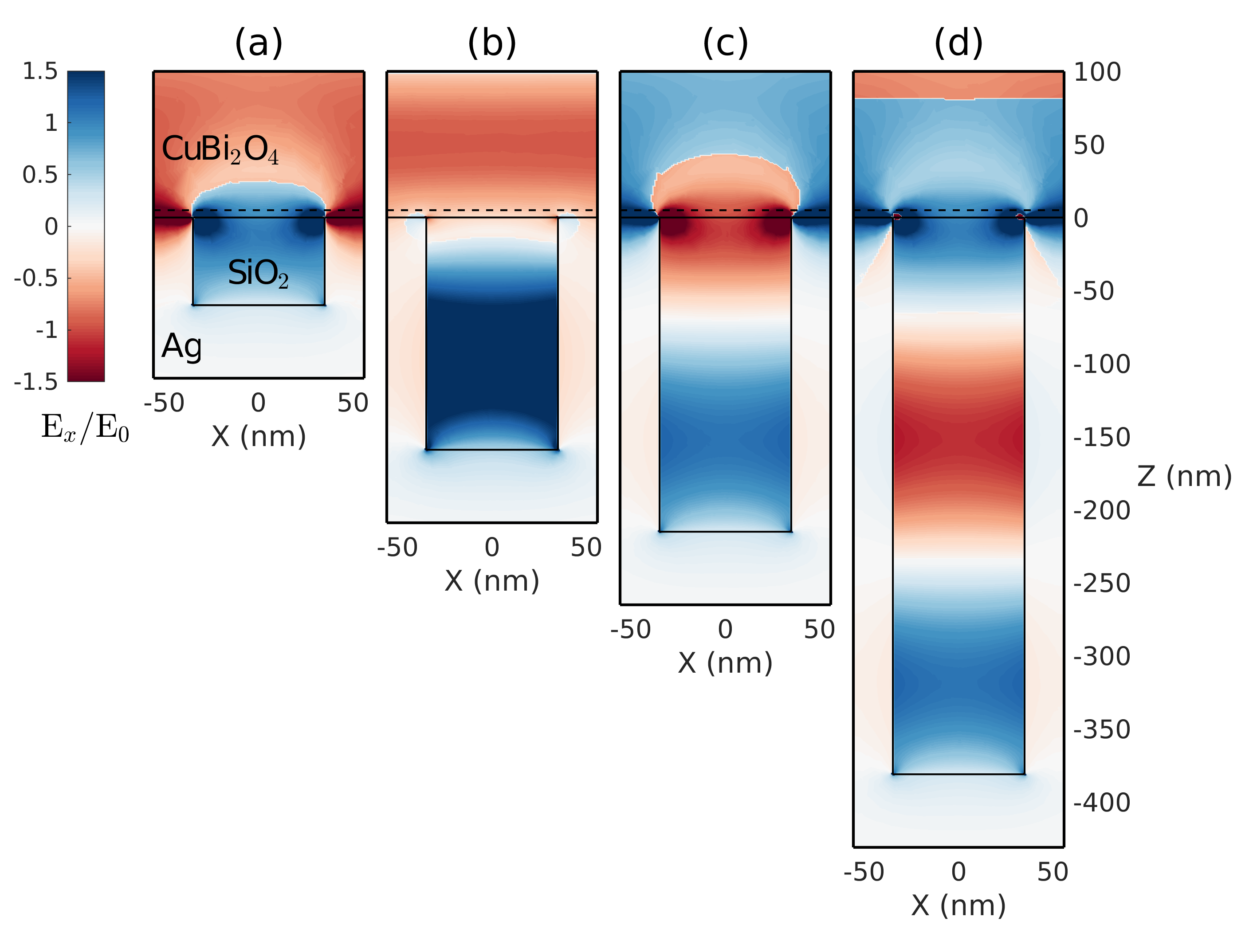}	
	\caption{The $x$ component of the electric field for the nanostructured 1D grating for slot lengths (a) 47~nm, (b) 159~nm, (c) 215~nm and (d) 381~nm. Light is incident normally from above with wavelength 650~nm and polarization in the $x-z$ plane.}
	\label{fig:Length_Variation_Near_Fields}
\end{figure}

Figure \ref{fig:Length_Variation} shows the variation in the absorption / reflection of the nanostructured device with respect to the nanoslot length at the four resonance wavelengths identified in figure \ref{fig:RTA}(b). In all four cases the absorption in both the semiconductor and the metal oscillates for increasing slot length. The $x$ component of the electric field for the four different slot lengths labeled in figure \ref{fig:Length_Variation}(d) is shown in figure \ref{fig:Length_Variation_Near_Fields}. These field patterns are typical for symmetric MIM modes \cite{main}{Kurokawa2007}, due to the field being localized entirely in the dielectric slot and oscillating with a wavelength close to the analytical MIM mode wavelength. The wavelength of the oscillation in the near field is also equal to the distance between the extrema in figure \ref{fig:Length_Variation}. The propagation constant $\beta_{MIM}$ of a MIM mode for an infinitely long dielectric slot surrounded by two semi-infinite half spaces of the same metallic material is given by the implicit set of equations \cite{main}{Maier2007},
\begin{eqnarray}
k_{1}^{2} = \beta_{MIM}^2-k_{0}^{2}\epsilon_{1}, \label{eq:MIM1} \\
k_{2}^{2} = \beta_{MIM}^2-k_{0}^{2}\epsilon_{2}, \\
\tanh \frac{k_{1}W}{2} = - \frac{k_{2}\epsilon_{1}}{k_{1}\epsilon_{2}}. \label{eq:MIM3}
\end{eqnarray}
where $k$ are wavenumbers in the given medium, $\epsilon$ are permittivities in each given medium and the subscripts 0,1 and 2 refer to vacuum, the dielectric and the metal, respectively. $W$ is the slot width. The wavelength of the mode can then be obtained from $\lambda_{MIM} = \Re(2\pi/\beta_{MIM})$.

In table \ref{tab:length_variation_wavelengths} the wavelength in the slot deduced from figures \ref{fig:Length_Variation} and \ref{fig:Length_Variation_Near_Fields} is compared to the MIM wavelength at the same vacuum wavelength for a slot width of 70~nm. The very close agreement between the two strongly suggests that these modes can be characterized as MIM modes and is in agreement with the analysis presented in \cite{main}{Wang2013}.

\begin{table}
	\centering
	\caption{Comparison of the theoretical metal-insulator-metal (MIM) plasmon resonance wavelength to the resonance wavelength of the slot modes. The MIM resonance wavelength was calculated assuming an infintely long Ag-SiO$_2$-Ag system with a SiO$_2$ thickness of 70~nm using equations \ref{eq:MIM1}-\ref{eq:MIM3}.}	
	\begin{tabular}{c | c c c c}		
		$\lambda_{vacuum}$ (nm) & 380 & 410 & 470 & 650 \\
		$\lambda_{MIM}$ (nm) & 128 & 168 & 216 & 324 \\
		$\lambda_{slot}$ (nm) & 156 & 189 & 231 & 344 \\
	\end{tabular}
	\label{tab:length_variation_wavelengths}
\end{table}

\begin{figure}
	\begin{center}
		\includegraphics[width=1.0\textwidth]{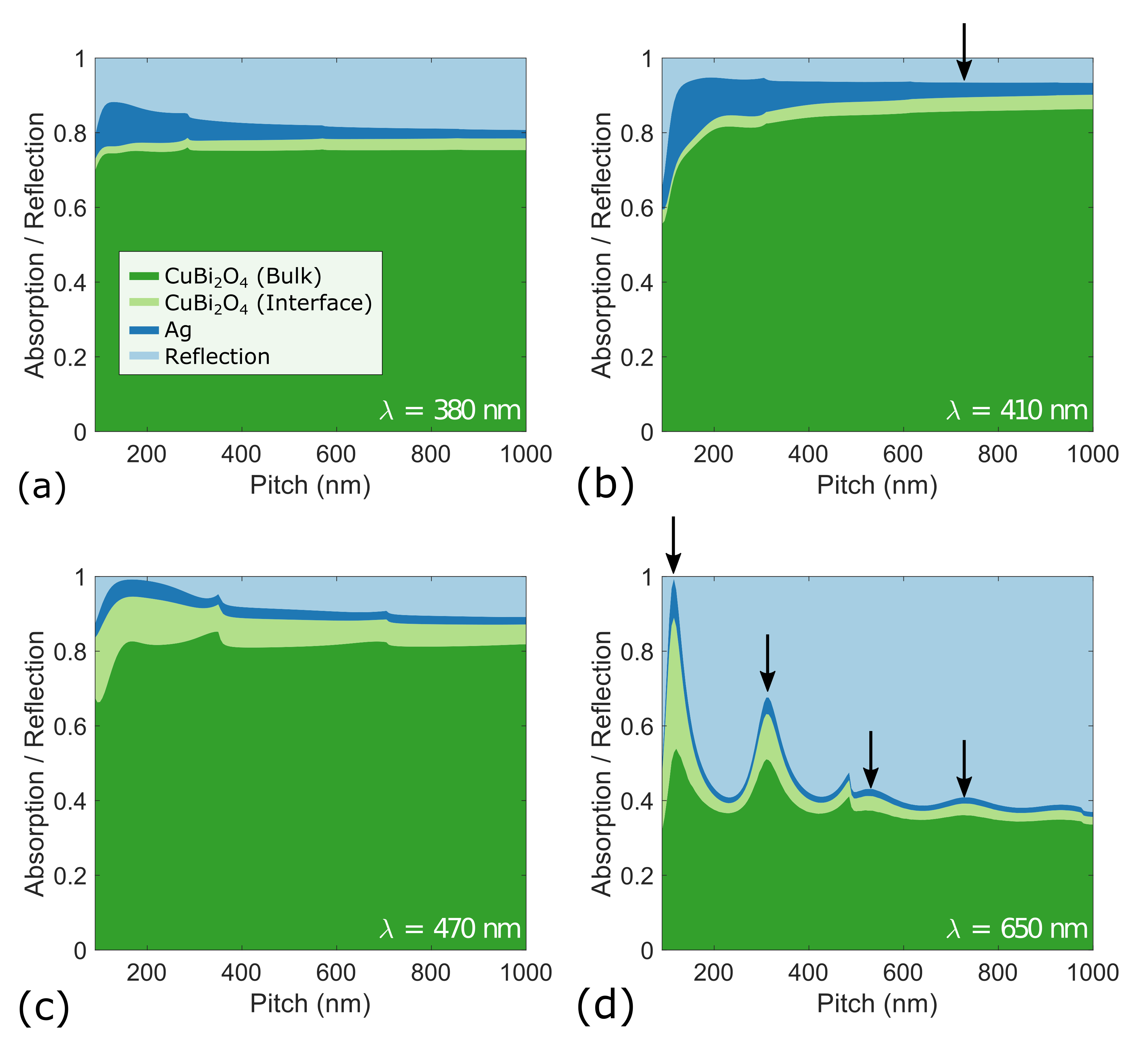}
	\end{center}	
	\caption{The absorption and reflection for the nanostructured 1D grating as a function of the grating pitch. The slot width and length are 70 and 60~nm, respectively. The wavelength of light is 380, 410, 470, 650~nm for (a-d), respectively. Arrows in (b,d) represent pitches where the electric field data is shown in figure \ref{fig:Pitch_Variation_Near_Fields}. Apart from the pitch, the geometry is the same as in the inset of figure \ref{fig:RTA}(b).}
	\label{fig:Pitch_Variation}
\end{figure}

Figure \ref{fig:Pitch_Variation} shows the variation in the absorption and reflection with respect to the grating pitch for the four resonance wavelengths labeled in figure \ref{fig:RTA}(b). For wavelengths 380 and 410~nm (a and b, respectively) there is very little dependence of the absorption and reflection on pitch. For very small pitches ($<$150~nm) the distance between the nanoslots is so small that there may be direct coupling between the modes in the slots, which can strongly affect the absorption and reflection. However for larger pitches the behavior is almost monotonic. The absorption in the metal decreases while the absorption in the semiconductor and the reflection increase. This is due to the slot having a smaller proportional effect for larger pitches. The fact that the modes do not show a strong dependence on pitch suggests that the MIM modes at these wavelengths couple to a localized mode since the only effect of changing the pitch is to monotonically reduce the interaction with the slot. It should be noted that Wood's anomalies are present at pitches of 285 and 305~nm for the wavelengths 380 and 410~nm, respectively. These arise due to the first diffraction order becoming available at these pitches \cite{app}{Enoch2012}.

\begin{figure}
	\centering
	\includegraphics[width=\textwidth]{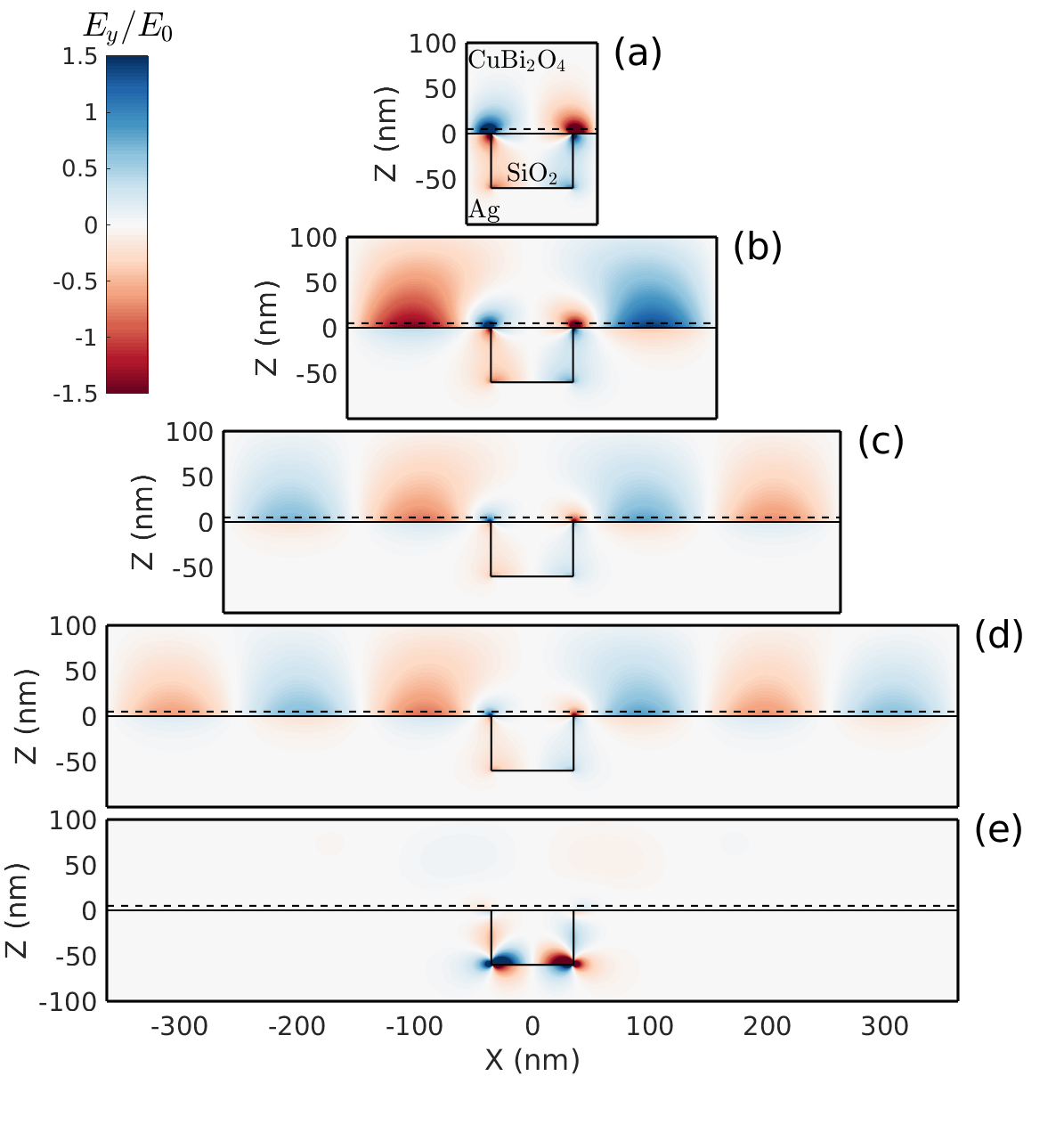}	
	\caption{The $y$ component of the electric field for the nanostructured 1D grating for grating pitches (a) 112~nm, (b) 315~nm, (c) 525~nm, (d) 725~nm and (e) 725~nm. (a-d) are for incident wavelength 650~nm while (e) is for incident wavelength 410~nm. Light is incident normally from above and polarized in the $x-z$ plane.}
	\label{fig:Pitch_Variation_Near_Fields}
\end{figure}

For the wavelengths of 470 and 650~nm the behavior is different. At $\lambda$ = 650~nm there are prominent oscillations in the absorption and in particular for the interface absorption with respect to the pitch. The first four peaks of this enhancement are explicitly labeled in (d). They occur at pitches of 112, 315, 525 and 725~nm. For the wavelength of 470~nm the oscillations in the interface are less obvious, nevertheless the first two peaks in the interface absorption are present at pitches of 105 and 205~nm. For both cases the distance between peaks is very close to the SPP wavelength for a SPP traveling at the CuBi$_2$O$_4$ / Ag interface.

The propagation constant ($\beta_{SPP}$) of a surface plasmon polariton at a planar interface was determined from \cite{main}{Maier2007},
\begin{equation}
\beta_{SPP} = k_{0}\sqrt{\frac{\epsilon_{1}\epsilon_{2}}{\epsilon_{1} + \epsilon_{2}}}.
\label{eq:SPP}
\end{equation}
Where $k_{0}$ is the wavenumber of light in a vacuum. $\epsilon_{1}$ and $\epsilon_{2}$ are the permittivities of Ag and CuBi$_2$O$_4$, respectively. The wavelength of the mode can then be obtained from $\lambda_{SPP} = \Re(2\pi/\beta_{SPP})$.

As in the previous case for the shorter wavelengths, the presence of Wood's anomalies can be see at pitches of 350 and 700~nm for 470~nm wavelength and pitches of 485 and 975~nm for 650~nm wavelength.

In order to further investigate the oscillations of absorption in figure \ref{fig:Pitch_Variation}(d) and the lack of any oscillation in figure \ref{fig:Pitch_Variation}(b), the electric field in the device for the pitches labeled by arrows in  \ref{fig:Pitch_Variation}(b,d) are show in figure \ref{fig:Pitch_Variation_Near_Fields}. For parts a-d the presence of a mode traveling at the CuBi$_2$O$_4$ / Ag interface is clearly visible. The distance between phase fronts in figure \ref{fig:Pitch_Variation_Near_Fields}a-d, which matches the distance between peaks in absorption enhancement in figure \ref{fig:Pitch_Variation}(d), is close to the SPP wavelength for an incident wavelength of 650~nm (see table \ref{tab:pitch_variation_wavelengths}). Therefore the interface contribution to absorption can be well characterized by a SPP mode with its source at the grating corners that is then absorbed mainly by the CuBi$_2$O$_4$. The peaks in enhancement occur when the distance between the grating corners and the periodic boundary (determined by the pitch) are equal to a half integer number of SPP wavelengths causing constructive interference.

\begin{table}
	\centering
	\caption{Comparison of the theoretical surface plasmon polariton (SPP) resonance wavelength  at a Ag / CuBi$_2$O$_4$ interface to the wavelength of the modes at the Ag / CuBi$_2$O$_4$ interface in the nanostructured solar fuel device. The SPP wavelength was calculated using equation \ref{eq:SPP}.}
	\begin{tabular}{c | c c c c}		
		$\lambda_{vacuum}$ (nm) & 380 & 410 & 470 & 650 \\
		$\lambda_{SPP}$ (nm) & 1147 & 490 & 205 & 112 \\
		$\lambda_{interface}$ (nm) & ~ & ~ & 200 & 100 \\
	\end{tabular}
	\label{tab:pitch_variation_wavelengths}
\end{table}

Figure \ref{fig:Pitch_Variation_Near_Fields}(e) shows the same pitch (725 nm) as in (d) but for the wavelength of 410~nm. In this case the resonance is clearly localized inside the slot. This agrees with the previous observation that the mode is independent of pitch variations. The localized mode is confined to the bottom corners of the grating, meaning no SPP can be emitted from the top corners.


\begin{figure}
	\begin{center}
		\includegraphics[width=0.5\textwidth]{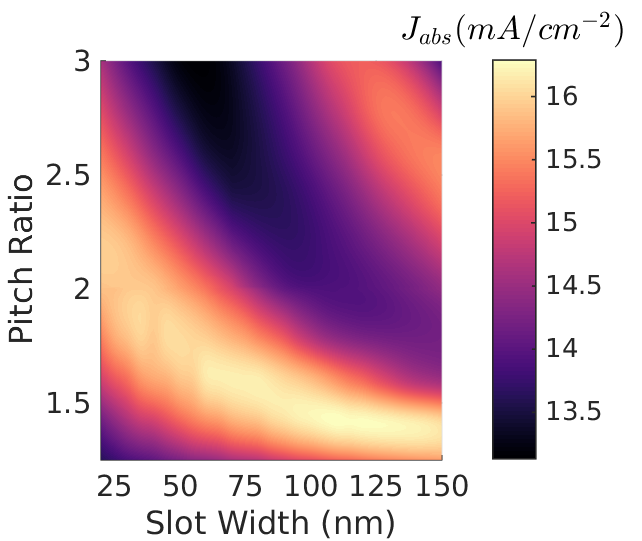}
	\end{center}
	
	\caption{The theoretical maximum current density ($J_{abs}$) achievable using the nanostructured back contact for different SiO$_{2}$ slot width and pitch ratio values. Values are for a 1D grating assuming p polarization.}
	\label{fig:Opt}
\end{figure}

\begin{figure}
	\begin{center}
		\includegraphics[width=1.0\textwidth]{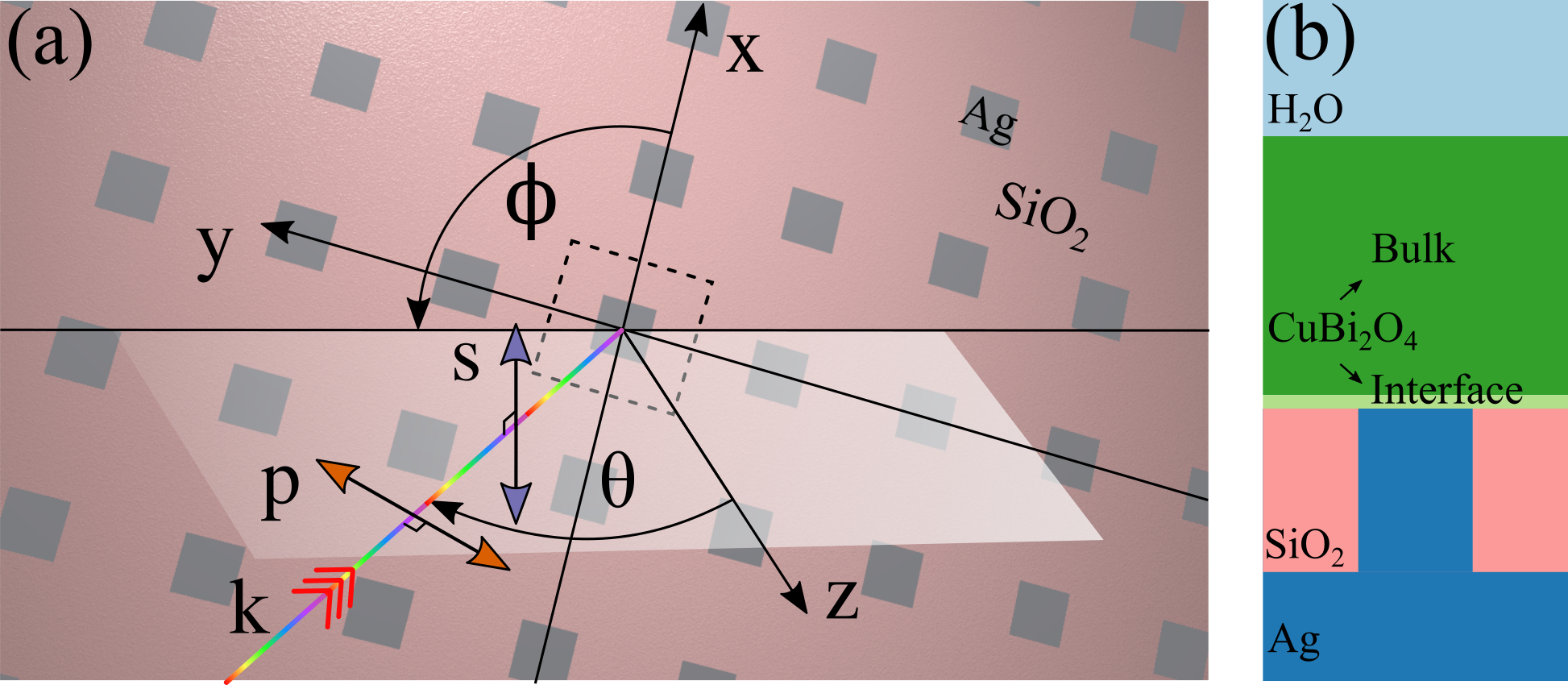}
	\end{center}
	
	\caption{(a) The 2D periodic nanostructure formed by Ag and SiO$_2$. The boundary of the periodic unit cell in the $x-y$ plane is shown by the dashed line. The azimuthal angle $\phi$ and polar angle $\theta$ define the direction of the $k$ vector of the incident plane wave and the scattering plane (white). The two linear polarizations which are parallel (p) and perpendicular (s) to the scattering plane are also shown. The angular resolved response of the full device including a 100~nm CuBi$_2$O$_4$ layer and upper H$_2$O and lower Ag half space is shown in figure \ref{fig:2DGrating}). (b) Cross section of the periodic unit cell of the 2D grating.}
	\label{fig:2DSchema}
\end{figure}

Figure \ref{fig:Opt} shows the maximum photocurrent density obtainable from the proposed slot waveguide back contact for different combinations of slot width ($W$) and pitch ratio $\rho = P/W$, where $P$ is the pitch. The value of $J_{abs}$ in each case has been taken using the optimal slot length found, therefore each position in figure~\ref{fig:Opt} has a slot length independent to those surrounding it. There are two main regions of absorption enhancement, the first is contained within the lower and left hand region of the parameter space. This corresponds to the resonance structure observed in fig~\ref{fig:RTA}(b) with two main absorption enhancement peaks between 470 and 650~nm. The geometrical parameters for this optimum value as given in the main text are: pitch = 112~nm, slot depth = 60~nm, slot width = 70~nm.

 As the slot width increases the distance between neighboring Ag regions increases, leading to the resonance at 650 nm from \ref{fig:RTA}(b) being redshifted out of the spectral region of interest. This can be counteracted somewhat by reducing the pitch, meaning that the region of maximum current density shifts to lower pitch ratios for higher slot widths. The second region of high current density is situated in the upper right region of figure~\ref{fig:Opt}. This region shows a larger number of resonances within the spectral region of interest due to higher order MIM modes arising from the larger pitch and slot width.

The region of maximum current enhancement is fulfilled with a broad range of slot width and pitch combinations, meaning that the proposed structure can be tailored to fit the constraints of the nanostructuring process used to create it.

Figure \ref{fig:2DSchema}(a) shows the coordinate system used for the 2D grating. Cuboid Ag nanoparticles periodically arrayed in the $x-y$ plane sitting on a Ag back reflector. The Ag nanoparticles are embedded in a SiO$_2$ matrix. The CuBi$_2$O$_4$ layer has not been shown for clarity. Also shown are the incident azimuthal angle ($\phi$) and polar angle ($\theta$) which define the $k$ vector of the incident light as well as the scattering plane. Depending on whether the electric field components are perpendicular or parallel to the scattering plane the incident light can be classified as s or p, respectively. Figure \ref{fig:2DSchema}(b) shows a cross section of the periodic unit cell of the 2D grating.

\bibliography{app}{references}{References}
\end{document}